\documentstyle[12pt]{article}
\title{\Large How Objective is Black Hole Entropy?}
\author{Y.K.Lau\\Institute of Theoretical Physics, Academia Sinica\\P.O.Box
2735, Beijing 100080\\ The People's Republic of China}

\date{}
\newcommand\eq{equation}

\begin{document}
\baselineskip=22 pt
\begin{titlepage}
\maketitle

\begin{abstract}
The objectivity of black hole entropy is discussed  in the particular case of a
Schwarzchild black hole.
Using Jaynes' maximum entropy formalism and Euclidean
path integral evaluation of partition
function, it is argued that
 in the semiclassical limit when the fluctutation of metric is neglected,
the black hole entropy of a Schwarzchild black hole is equal to the maximal
information entropy of an observer whose sole knowledge of the black hole is
its mass. Black hole entropy becomes
 a measure of number of its internal mass eigenstates in accordance with the
Boltzmann principle only in the limit of negligible relative mass fluctutation.
{}From the information theoretic perspective, the example of
 a Schwarzchild black hole seems to suggest that black hole entropy is no
different from ordinary thermodynamic entropy. It is a property of the
experimental data of a black hole, rather than being an intrinsic physical
property of a black hole itself independent of any observer. However, it is
still weakly objective in the sense that different observers given the same set
of data of a black hole will measure the same maximal information entropy.
\end{abstract}
 \end{titlepage}
Given a black hole with certain mass, charge and momentum. From the
consideration of particle creation of  a black hole${}^1$, it may be assigned
an entropy like quantity given by the Bekenstein Hawking formula as
$$S_{BH}={1\over 4}A$$
where $A$ is the area of the event horizon. The units in which $G=\hbar=c=k=1$
will be used throughout
where all symbols have their usual meanings. $S_{BH}$ is usually associated
with the loss of information of the microstates of a black hole for an observer
outside the event horizon who, according to the no hair theorem, knows only the
mass, charge and angular momentum of the hole${}^2$. However, the association
of $S_{BH}$ with loss of information remains at  a heuristic level so far,
nothing more than as a guide of thinking.

On the other hand, black hole entropy may be identified up to a constant with
the area of the event horizon. Since the area of the event horizon is geometric
in character, it does not seem to depend on any coarse graining process or the
presence of an observer.
This led to the speculation ${}^4$ that black hole entropy is more
fundamental  than its counterparts for other physical systems. It is an
objective,
intrinsic physical attribute of a black hole, in equal footing with the mass,
charge and angular momentum.

The aim of this note is to examine these two problems
in the particular case of a Schwarzchild black hole, in the hope that this will
shed some light on the issues in general. We shall show that
 in the semiclassical limit when the
fluctutation of the metric is negligible,
\par\noindent
(i) Black hole entropy given by the Bekenstein Hawking formula is indeed equal
to the maximal information entropy  (in a sense to be explained below) of an
observer whose sole knowledge of the black hole is its mass.
In doing so, we also see that
black hole entropy is a measure of the number of internal mass eigenstates of a
Schwarzchild black hole, in accordance with the Boltzmann principle only in the
limit of negligible relative fluctutation of mass.
\par\noindent
(ii) The claim of objectivity in the sense described above of black hole
entropy does not seem to stand. However, a weaker form of objectivity is still
preserved

\section*{\sl Information Entropy of Schwarzchild Black Hole}
Consider a Schwarzchild black hole with mass $M$. From a thermodynamic
perspective, $M$ is the average mass of the black hole. The real mass
fluctutates around $M$. To simplify discussion, assume the black hole has
discrete mass spectrum
and denote its mass eigenstates by the mass eigenvalues $M_i>0,i=1\cdots N$ for
some positive integer $N$. The consideration of continuous mass spectrum is
similiar but mathematically more complicated. It does not seem to add any
further physical insight in the present context.

Let $p_i$ be the probability that the black hole is in the $M_i$ state subject
to the constraint that $\sum p_iM_i=M$ and $\sum p_i=1$. Throughout summation
sign is understood to be a sum from $i=1$ to $N$. Given only $M$, the
assignment of $p_i$ is quite arbitrary. However, if we invoke the Jaynes'
principle${}^{5,6}$, then the most unbiased assignment would be $p_i$ which
maximises the quantity
$$\sum p_i\ln p_i+\lambda M+\mu\sum p_i$$
where $\lambda,\mu$ are Lagrange multiplers. This gives
\begin{\eq}
\label{e1}
p_i={1\over Z}e^{-\lambda M_i}
\end {\eq}
for $i=1\cdots N$. $Z$ is the partition function given as
\begin{\eq}
\label{e2}
Z=\sum e^{-\lambda M_i}
\end {\eq}
In addition, we have
\begin{\eq}
\label{e3}
M=-{{\partial\ln Z}\over{\partial\lambda }}
\end {\eq}
\begin{\eq}
\label{e4}
S_I=\lambda M+\ln Z
\end {\eq}
and
\begin{\eq}
\label{e4a}
\lambda=M={{\partial {S_I}}\over{\partial M}}
\end {\eq}
where $S_I=-\sum p_i\ln p_i$ is the maximal information entropy. The word
maxiaml is used in the sense that for any other arbitrary assignment of $p_i'$
to $M_i$ subject to the same constraint $\sum p_i'M_i=M$ and $\sum p_i'=1$, the
information entropy $-\sum p_i'\ln p_i'$ is less than or equal to $S_I$.
With $M_i>0$ for $i=1,2\cdots N$, it may be shown from ~(\ref{e4a}) that
$\lambda>0$.

So far, the above argument is true for any system  characterised by a parameter
whose parameter has mean value $M$.
To proceed, we need to evaluate further the partition function in ~(\ref{e2}).
Note that ~(\ref{e2}) may be expressed as a path integral ${}^7$
\begin{\eq}
\label{e5}
\int e^{iI} Dg
\end {\eq}
where the functional integral is over all Euclidean, asymptotically flat
metrics whose imaginary time variables have period $\lambda$. $I$ is the action
for a metric $g$  given as
$$ I=  -{1\over{16\pi}}\int_M R\,\,+{1\over{8\pi}}\int_{\partial M}[K]$$
Here $R$ is the Ricci scalar, $M$ is a compact region in an asymptotically flat
Riemannian manifold bounded by the boundary $\partial M$ which is topologically
$S^1\times S^2$. $[K]$ is the difference between the trace of the second
fundametal form of $M$ in $g$ and the Euclidean flat space metric.

Observe that among all Euclidean, asymptotically flat metrics satisfying the
prescribed periodic boundary condition, the classical action is given
by the Euclidean Schwarzchild metric with mass ${\lambda\over{8\pi }}$. Since
the path integral in ~(\ref{e5}) is dominated by contribution from the
classical action, using the stationary phase approximation and neglecting
fluctutation of the metric around the classical background, one may readily
show that ${}^7$
\begin{\eq}
\label{e6}
\ln Z=-{{\lambda^2}\over{16\pi}}
\end {\eq}
{}~(\ref{e3}) and ~(\ref{e6}) together imply
\begin{\eq}
\label{e7}
\lambda=8\pi M
\end {\eq}
{}From ~(\ref{e4}) and ~(\ref{e7}), we may then infer
\begin{\eq}
\label{e8}
S_I=S_{BH}
\end {\eq}
The path integral evaluation of partition function is essentially due to
Gibbons and Hawking${}^7$. However, unlike that in their work, the classical
black hole metric concerned is not a priori included in the path integral.
Instead, the relation between the path integal and the black hole concerned is
given by (3). In doing so, we may see  more clearly the connection between
information entropy in statistical mechanics and black hole entropy.

\section*{\sl Boltzmann principle}
Black hole entropy is a measure of number of microstates of a black hole was
first
 conjectured  by Bekenstein${}^2$. It was later confirmed in an affirmative way
by Zurek${}^3$. Here if we write (1)
\begin{\eq}
\label{e9}
p_i={1\over Z}e^{-\lambda M(1+{{\Delta M_i}\over M})}
\end{\eq}
where $\Delta M_i=M_i-M$. It is clear that in the limit the relative
fluctutation of mass ${{\Delta M_i}\over M}\rightarrow 0$ for $i=1\cdots N$,
we have from ~(\ref{e9}), ~(\ref{e4}) and ~(\ref{e8})  that
\begin{\eq}
\label{e10}
S_{BH}=S_I=\ln N
\end{\eq}
in accordance with the Boltzmann principle. In the present context, apart from
seeing ~(\ref{e10}), it also shows that Boltzmann formula is a limit expression
for the black hole entropy when the thermal fluctutation of mass is negligible
compared with the mass itself, as it should be the case for the thermodynamic
description of a physical system to be valid.

However, one also has to admit that
the above argument is not entirely satisfactory since it does not involve the
thermodynamic limit $N\rightarrow \infty$. One would like a more rigorous limit
theorem which asserts that, subject to some restriction on the variance on
$\Delta M_i$, ${{\Delta M_i}\over M}\rightarrow 0$ as $N\rightarrow \infty$ for
$i=1\cdots N$. However, due to the negative heat capacity for a Schwarzchild
black hole, the canonical ensemble constructed by Jaynes' variational principle
is not stable under fluctutation of $M$. This manifests in the absurdity that
the
variance of $\Delta M_i$ is negative. At the same time, the mean of $\Delta
M_i$
is zero. This renders the formulation of such a theorem along the line of the
law of large numbers difficult.

\section*{\sl How Objective is Black Hole Entropy?}
Ever since the inception of black hole entropy into physics, due to its
geometric character, there has been speculation${}^4$ that black hole entropy
is more fundamental than its counterparts for other physical systems. It is an
intrinsic, objective physical property of a black hole, in equal footing with
for example the mass.
 We shall examine this belief in what follows in view of the equality between
$S_I$ and $S_{BH}$ established in ~(\ref{e8}).

{}From the perspective of Jaynes' maximum entropy formalism, equilibrium
thermodynamic entropy is merely a special case of information entropy, being
the maximal information entropy subject to a given set of time independent
experimental data. Therefore
equilibrium thermodynamic
 entropy is defined relative  to a set of experimental data, rather than being
a physical characteristic of a system, like energy, pressure etc. This view was
first put forward by Jaynes ${}^8$ and subsequently given careful
linguistic formulation by Denbigh ${}^9$. Different set of experimental data
will give rise to different maximal information entropy. Even so, the notion of
 equilibrium thermodynamic entropy still preserves weak
objectivity${}^{8,9}$in the sense that different observers given the same set
of experimental data will agree on the
maximal information entropy assigned to the data. From this information
 theoretic perspective, in view of ~(\ref{e8}), black hole entropy  of a
Schwarzchild black hole
is no different  from the  equilibrium thermodynamic entropy of other
thermodynamical systems. It is defined relative to the given mass data of the
black hole.
Being able to identify $S_I$ with a geometric quantity in spacetime should be
looked on as merely an aesthetically appealing coincidence.
This explains why an observer travelling beyond  the event horizon into a black
hole will not measure any black hole entropy${}^1$. Depending on the
information in hand about the microstates of a black hole, an observer inside
the event horizon will assign a  maximal information entropy in accordance with
the data in hand. This maximal information entropy will in general be different
from that of an observer who only knows the mass of the black hole.
 In view of the different assignment of maximal information entropy to the same
black hole for observers inside and outside the event horizon, it seems that,
contrary to many would like to believe, black hole entropy of a Schwarzchild
black hole
provides another example which supports Jaynes' thesis that thermodynamic
entropy is not a physical property of a thermodynamic system, it is  defined
relative to a set of experimental data.

\section*{\sl Concluding Remarks}
We have shown that, in the particular case of a Schwarzchild black hole, in the
semiclassical limit, the black hole entropy may be identified with the maximal
information entropy of an observer who knows only the mass of the black hole.
Various implications of this result are also discussed. In particular, it
throws considerable doubt that entropy will acquire an objective status,
similiar to other physical attributes of a physical system in the context of
black hole physics. It remains to investigate whether in the generic case of a
Kerr Newman black hole, the maximal information entropy of an observer who
knows only the mass, charge and angular momentum of the black hole may be
identified with the Bekenstein Hawking black hole entropy. It seems that more
sophiscated argument than that presented here is needed to achieve this general
result. The difficulty lies not so much in the evaluation of the partition
function but in the fact that, instead of (3), we have a system of PDE whose
solution is sought for.

\vspace{40pt}
\par\noindent
{References}
\par\noindent
${}^1$ S.W. Hawking, Comm. Math. Phys. {\bf 43} 199 (1975)
\par\noindent
${}^2$ J.D. Bekenstein, Phys. Rev. D. {\bf 7} 2333 (1973)
\par\noindent
${}^3$ W.H. Zurek, Phys. Rev. Lett. {\bf 50} (13) 1013 (1983); W.H. Zurek and
K. Thorne., Phys. Rev. Lett. {\bf 54} 2171 (1985)
\par\noindent
${}^4$ N.D. Birrell and P.C.W. Davies., {\sl Quantum Fields in Curved Space}
(CUP, 1982), section 8.5; R. Penrose, {\sl Singularities and time asymmetry},
section 12.4 in {\sl General Relativity, an Einstein centenary survey} ed. by
S.W. Hawking and W. Isarel (CUP,1979)
\par\noindent
${}^5$ E.T. Jaynes, {\sl Papers in Probability, Statistics and Statistical
Physics}
\par\noindent
  Rosenkrantz (ed.) (Reidel Dordrecht, 1983)
\par\noindent
${}^6$ Jaynes' principle was first used by Bekinstein in the context of black
holes. See J.D. Bekenstein, Phys. Rev.D 12(10) 3077(1975).
\par\noindent
${}^7$ G.W. Gibbons and S.W. Hawking., Phys. Rev. D {\bf 15} (10) 2752 (1977)
\par\noindent
${}^8$ E.T. Jaynes, Amer. J. of Phys. {\bf 33} 391 (1965); also reprinted in
(5).
\par\noindent
${}^9$ K. Denbigh, Chem. Brit. {\bf 17} 168 (1981); also reprinted in H.S.Leff
and A.F.Rex (ed.) {\sl Entropy, Information and Computing} (Adam Hilger,
Bristol, 1990)

\end{document}